\documentstyle[12pt,twoside]{article}

\catcode`\@=11
\def\newsymbol#1#2#3#4#5{\let\next@\relax%
 \ifnum#2=\@ne\else%
 \ifnum#2=\tw@\let\next@\msyfam@\fi\fi%
 \mathchardef#1="#3\next@#4#5}
\def\mathhexbox@#1#2#3{\relax%
 \ifmmode\mathpalette{}{\m@th\mathchar"#1#2#3}
 \else\leavevmode\hbox{$\m@th\mathchar"#1#2#3$}\fi}
\def\hexnumber@#1{\ifcase#1 0\or 1\or 2\or 3\or 4\or 5\or 6\or 7\or 8%
\or 9\or A\or B\or C\or D\or E\or F\fi}
\font\tenmsy=msbm10
\font\sevenmsy=msbm7
\font\fivemsy=msbm5
\newfam\msyfam
\textfont\msyfam=\tenmsy
\scriptfont\msyfam=\sevenmsy
\scriptscriptfont\msyfam=\fivemsy
\edef\msyfam@{\hexnumber@\msyfam}
\def\Bbb#1{\fam\msyfam\relax#1}
\catcode`\@=\active
\load{\footnotesize}{\sf}

\newtheorem{theorem}{Theorem}[section]
\newtheorem{proposition}[theorem]{Proposition}
\newtheorem{lemma}[theorem]{Lemma}
\newtheorem{corollary}[theorem]{Corollary}
\newtheorem{definition}[theorem]{Definition}
\newtheorem{example}[theorem]{Example}
\newtheorem{remark}[theorem]{Remark}

\newcommand{\X}{{\Delta^\dagger_t}}
\newcommand{\XX}{{\Delta^\dagger_{K'_t}}}
\newcommand{\XY}{{\Delta^\dagger_{K_t}}}

\newcommand{\Y}{\Delta^\dagger}

\newcommand{\TT}{( e^{t\dA})_{21}}
\renewcommand{\Im}{{\rm Im}}
\renewcommand{\Re}{{\rm Re}}

\newcommand{\SS}{\sigma_2}
\newcommand{\OO}{(e^{t\dA})_{11}} 
\newcommand{\KK}{\frac{K_t}{t}} 
\newcommand{\K}{{K_t}}

\newcommand{\TA}{\tau_{\dA}}

\newcommand{\ea}{e^{t\dA}}

\newcommand{\fm}{\phi_m}
\newcommand{\fn}{\phi_n}
 
\newcommand{\sn}{\psi_n}

\newcommand{\bfn}{\ovv{\phi}_n}
\newcommand{\bsm}{\ovv{\psi}_m} 
\newcommand{\bsn}{\ovv{\psi}_n}
\newcommand{\T}[1]{\ov{#1^\ast}}

\newcommand{\lk}{\left(}

\newcommand{\rk}{\right)}
\newcommand{\lkk}{\left\{}
\newcommand{\lkkk}{\left[}
\newcommand{\rkkk}{\right]}

\newcommand{\rkk}{\right\}}
\newcommand{\mat}[4]{\lk\!\!\begin{array}{cc}
#1&#2\\
#3&#4
\end{array}
\!\!\rk}
\newcommand{\ov}[1]{\overline{#1}} 
\newcommand{\ovv}[1]{\ov{#1}} 
\newcommand{\oa}[1]{\ov {{#1}^\ast}}
\newcommand{\ST}{\mat{S}{\ov T}{T}{\ov S}}

\newcommand{\U}{U}
\newcommand{\UB}{U_{\rm B}}

\newcommand{\limn}{\lim_{n\rightarrow\infty}}

\newcommand{\dA}{\widehat{A}}

\newcommand{\limM}{\!\!\lim_{M\rightarrow\infty}}

\newcommand{\limt}{\lim_{t\rightarrow0}}

\newcommand{\proof}{{\noindent \it Proof:\ }}
\newcommand{\qed}{\hfill $\Box$\par\medskip}
\newcommand{\BR}{{{\Bbb R}^{d}}}
\newcommand{\D}{\Delta}

\newcommand{\ii}{\infty}

\newcommand{\BC}{{{\Bbb C}}}

\newcommand{\RR}{{\Bbb R}}

\newcommand{\fff}{{{\cal F}}}
\newcommand{\A}{A}

\newcommand{\ffff}{{{\cal F}_0}}
\newcommand{\B}{{\rm B}}
\newcommand{\BT}{{{B_\oplus^2}}}

\newcommand{\cnk}{\lk\!\!\begin{array}{c}n\\k\end{array}\!\!\rk}
\newcommand{\ia}{i\D(\dA )}

\newcommand{\HS}{B_2} 

\newcommand{\Tr}{{\rm Tr}}
\newcommand{\I}{1} 
\newcommand{\ui}[1]{e^{-\half \D_{#1}^\dagger}}
\newcommand{\uii}[1] {\wick{e^{- N_{#1}}}}
\newcommand{\uiii}[1]{e^{-\half \D_{#1}}}

\newcommand{\intt}{\int_0^t\int_0^{t_1}\cdots \int_0^{t_{n-1}}} 
\newcommand{\dttt}{dt_1dt_2\cdots dt_n} 
\newcommand{\dd}[1]{\D^\dagger_{#1}}
\newcommand{\da}[1]{N_{#1}}
\renewcommand{\aa}[1]{\D_{#1}}

\newcommand{\ad}{{\rm ad}}

\newcommand{\wick}[1]{:\!\!{#1}\!\!:}
\newcommand{\f}{^{-1}}
\newcommand{\sumn}{\sum_{n=0}^\ii}
\newcommand{\sumnn}{\sum_{n=1}^M}
\newcommand{\sumnnm}{\sum_{n=0}^M}

\newcommand{\half}{{\frac{1}{2}}}

\newcommand{\bdd}{b_\A^\ast}
\newcommand{\bbb}{b_\A} 

\newcommand{\ddd}{{\cal D}_\infty}
\newcommand{\add}{a^\ast}
\newcommand{\bss}{b_\A^\sharp}
\newcommand{\ass}{a^{\sharp}}

\newcommand{\han}{{1/2}}

\newcommand{\hhh}{{\cal H}}
\newcommand{\hhhh}{{\hhh \oplus\hhh}} 

\newcommand{\la}{{\lambda}}

\newcommand{\eqq}[1]{\label{#1}}
\newcommand{\eq}[1]{\begin{equation}
\label{#1}}
\newcommand{\en}{\end{equation}}

\newcommand{\be}{\begin{eqnarray*}}
\newcommand{\ee}{\end{eqnarray*}}
\newcommand{\bee}{\begin{eqnarray}}
\newcommand{\eee}{\end{eqnarray}}
\newcommand{\nn}{\nonumber}

\newcommand{\kak}[1]{(\ref{#1})}

\newcommand{\bt}[1]{\begin{theorem}
\label{#1}}
\newcommand{\et}{\end{theorem}}
\newcommand{\bl}[1]{\begin{lemma}
\label{#1}}
\newcommand{\el}{\end{lemma}}
\newcommand{\bp}[1]{\begin{proposition}
\label{#1}}
\newcommand{\ep}{\end{proposition}}
\newcommand{\bi}{\begin{itemize}}
\newcommand{\ei}{\end{itemize}}
\newcommand{\bc}[1]{\begin{corollary}
\label{#1}}
\newcommand{\ec}{\end{corollary}}

\newcommand{\bog}{\sp_2}
\renewcommand{\sp}{\Sigma}

\newcommand{\Sd}{\mat S 0 0 {\ov S}}
\newcommand{\Sdd}{{\bf S}} 
\newcommand{\Td}{\mat 0 {\ov T} {T} 0} 
\newcommand{\Tdd}{{\bf T}} 
\newcommand{\bT}{{\bf T}}

\newcommand{\dt}[1]{\frac{d}{d(t^2)}{#1}\lceil_{t=0}}
\newcommand{\dtt}[1]{\frac{d}{dt }{#1}\lceil_{t=0}}
\newcommand{\tu}{\frac{U_t-1}{t}}
\renewcommand{\det}{{\rm det}(1-\K^\ast \K)^{1/4}}

\newcommand{\rr}{\right\|}
\renewcommand{\ll}{\left\|}

\newcommand{\bb}{b^\ast_t}

\makeatletter
\@addtoreset{equation}{section}
\makeatother

\begin{document}
\setlength{\baselineskip}{12pt}
\title
{Local exponents and infinitesimal generators of 
 canonical transformations on Boson Fock spaces}
\author{K. R. Ito\thanks{e-mail:  ito@mpg.setsunan.ac.jp} and 
F. Hiroshima\thanks{e-mail:  hiroshima@mpg.setsunan.ac.jp}}

\date{\today}
\maketitle
\begin{abstract}
A one-parameter symplectic group 
$\{e^{t\dA}\}_{t\in\RR}$ derives 
 proper canonical  transformations on a Boson Fock space.  
It has been known that the unitary operator $U_t$ 
implementing such a proper canonical 
transformation 
gives a projective unitary representation of $\{e^{t\dA}\}_{t\in\RR}$ 
and that $U_t$ 
can be expressed as a normal-ordered form. 
We rigorously derive the self-adjoint operator $\D(\dA)$ and 
a phase factor 
$e^{i\int_0^t\TA(s)ds}$ with a real-valued function $\TA$ 
such that 
$U_t=e^{i\int_0^t\TA(s)ds}e^{it\D(\dA)}$. 
\end{abstract}

{\footnotesize 
\setlength{\baselineskip}{12pt}
}
\setlength{\baselineskip}{20pt}

\section{Introduction}
This paper is motivated by a close resemblance 
between 
infinitesimal generators of rotation groups in  {\it white noise analysis} and 
those  of {\it proper canonical  transformations} 
(Bogoliubov transformations)
on a Boson Fock space $\fff$. 
In white noise analysis rotation groups $\{g_\theta\}_{\theta\in\RR}$ 
acting on $(S')$ have been studied so far by many authors, e.g., see Hida
 \cite{hi}.  
Here $(S')$ is the dual of a subspace $(S)$ of  $\fff$. 
Such rotation groups are induced from e.g., 
shifts, dilations, $SO(n)$,  special conformal transformations  and 
the L\'evy group, etc. 
Their infinitesimal generators are defined by 
$$\lim_{\theta\rightarrow 0}  \frac{g_\theta-1}{\theta}\ \ \ in \ (S')$$ 
and 
it is established that generators 
consist of infinite dimensional Laplacians, e.g., 
the Gross Laplacian,  the L\'evy Laplacian, the Beltrami Laplacian, etc. 
These Laplacians are expressed 
in the form of polynomials of second degree of 
the annihilation operators and  the creation operators in $(S')$. 
See e.g. Obata \cite{o2}.

Meanwhile, 
it is known that  a proper canonical  transformation 
$$\UB(A) :\fff\rightarrow\fff,\ \ \ A\in\bog,$$    
gives a {\it projective} unitary representation of 
a symplectic group $\bog$ (\cite{sh}). 
In this paper from $\UB(\cdot)$ a unitary representation $\widehat U_t:=e^{-i\theta(t)}\UB(e^{t\dA})$ of  
a one-parameter symplectic group $\{e^{t\dA }\}_{t\in\RR}\subset\bog$ is constructed
with some real-valued  function $\theta(t)$.
Classically this is known as the Bargmann theorem \cite{ba}.
The  unitary representation $\widehat U_t$ 
induces a one-parameter unitary group $\{\widehat U_t\}_{t\in\RR}$ 
on $\fff$. 
Regarding $\fff$ as $\fff\subset (S')$, we recognize that 
the   self-adjoint infinitesimal generator $\D(\dA)$ of $\widehat U_t$  
 may correspond to  a Laplacian in $\fff$. 
The purpose of this paper is to give 
 $\D(\dA)$ and $\theta(t)$  {\it rigorously}.
\begin{figure}[ht]
\unitlength 5mm 
\begin{picture}(10, 8)(-17,-16) 
\thicklines
\put(-6,-10){$\fff$}
\put(0,-10){$(S')$}
\put(-3,-10){$\subset$}
\put(-5.5,-10.5){\vector(0,-1){3.5}}
\put(-7.5,-12.5){$\widehat U_t$} 

\put(-6,-15){$\fff$}
\put(0,-15){$(S')$}
\put(-3,-15){$\subset$}
\put(0.5,-10.5){\vector(0,-1){3.5}}
\put(1,-12.5){$g_\theta$} 
\end{picture}
\caption{One-parameter unitary group $\widehat U_t$ and rotation group $g_\theta$}
\label{hijk}
\end{figure}
\\ 
Let $\hhh$ be a  Hilbert space over the complex field $\BC$ 
endowed with an antiunitary involution $\Gamma$ with $\Gamma^2=1$.
Set  $$\ov X=\Gamma X \Gamma$$ 
for an operator acting on ${\cal H}$.  
Let $\sp$ be a symplectic group, i.e., 
$A=\ST\in\sp$ if and only if bounded operators $S,T$ on a Hilbert space $\hhh$ 
satisfy 
$$A^\ast J A=A J A^\ast =J,$$
where 
$$J=\mat 1 0 0 {-1}.$$
The linear canonical  transformation on  $\fff$ 
over $\hhh$ 
associated with  $A=\ST\in \sp$ is given as the map 
\be
& & 
 a(f)\longmapsto \bbb(f):=a(Sf)+\add(Tf)\\
& & 
 \add(f)\longmapsto \bdd(f):= a(\ov Tf)+\add(\ov Sf),
\ee 
where $\add(f)$ and $a(f)$, $f\in\hhh$,  are the creation operator and 
the annihilation operator on $\fff$  
smeared by $f$, respectively.
Let us define subgroup $\sp_2$  by 
$$\sp_2=\lkk \left.A=\ST\in\sp\right|T  \mbox{ a Hilbert-Schmidt operator}\rkk.$$
It had been established that the  linear canonical     transformation 
associated with $A\in\sp$ 
can be implemented by a unitary operator $\UB(A)$ on $\fff$  
if and only if $A\in\sp_2$, i.e., 
\eq{pr}
A\in\sp_2\Leftrightarrow 
\UB(A)\ass(f) \UB(A)\f =\bss (f),
\en 
where 
$\ass=a, \add$ and $\bss=b_A,\bdd$. 
 The correspondence 
between $a, \add$ and $b_A, \bdd$ arising as in \kak{pr}  is called 
a proper canonical transformation. 
Without effort one can show that if ${\rm dim }\hhh<\infty$, 
all  linear canonical transformations are proper.  
It is known that 
$\UB(A)$ is equal to 
\eq{uau}
\U(\A):={\rm det}(\I-K_1^\ast K_1)^{1/4}
\ui{K_1}{\uii{K_2}}\uiii{K_3} 
\en 
{\it up to a phase factor}. 
Here 
$\wick{X}$ denotes the normal ordering of $X$, 
$$K_1=TS\f,\ \ \ K_2=1-\ov{(S\f)^\ast},\ \ \ K_3=-S\f \ov T,$$
and 
the operators $\dd{X}$, $\aa X$, $\da X$ 
are   quadratic operators of $\add\add$ type, $\add a$ type and $aa$ type respectively,
which are  constructed by operator $X$. 
$U(A)$  is called 
the ``normal-ordered quadratic expression'' of $\UB(A)$. 
$$U(\cdot):\bog\longrightarrow {\rm unitary\  operators\ on \ \fff}$$
induces  a projective unitary representation 
of  $\sp_2$, i.e., 
$$U(A)U(B)=\omega(A, B)U(AB),\ \ \ A, B\in\sp_2,$$  
with local factor  $\omega(A, B)\in\{e^{i\psi}| \psi\in\RR\}$. 
We are interested in one-parameter symplectic groups in $\bog$. 
Let 
$$\SS:=\lkk \dA=\ST\left| 
\dA J+J\dA=\dA^\ast J+J\dA^\ast=0, T \mbox{ a Hilbert-Schmidt operator}\right.\rkk.
$$A one-parameter symplectic group $\{e^{t\dA }\}_{t\in\RR}$ 
belongs to $\bog$ if and only if $\dA\in\SS$.
When $\dA\in\SS$, 
$$e^{t\dA }=\mat{(e^{t\dA })_{11}}{(e^{t\dA })_{12}}{(e^{t\dA })_{21}}{(e^{t\dA })_{22}}
$$ 
induces the proper  canonical transformation 
\be
&& a(f)\longmapsto b_t(f):=a((e^{t\dA })_{11}f) +\add((e^{t\dA })_{21}f),\\
&& \add (f)\longmapsto b_t^\ast(f):=a((e^{t\dA })_{12}f) +\add((e^{t\dA })_{22}f).
\ee 
Set 
\be 
&&U_t:=U(e^{t\dA }),\ \ \ t\in\RR,\\
&&e^{i\rho(t,s)}:=\omega(e^{t\dA}, e^{s\dA}),\ \ \ t,s\in\RR,
\ee 
where $\rho(t,s)$ is called a local exponent. 
Then $U_t$ satisfies 
$$U_tU_s=e^{i\rho(t,s)} U_{t+s},\ \ \ t,s\in\RR.$$
In this paper we shall show that 
\bee
\eqq{q}
&& \widehat U_t:=
e^{-i\int_0^t \TA(s)ds}
 U_t,\\
&& \TA(s) :=\half \Im {\rm Tr}(T^\ast (e^{s\dA})_{21}(e^{s\dA})_{11}\f),\nn
\eee satisfies 
$$
\widehat U_t\widehat U_s=\widehat U_{t+s}.
$$
Namely local exponent $\rho(t,s)$ is equivalent to zero, i.e., 
$\widehat U_t$ furnishes a unitary representation of $\{e^{t\dA}\}_{t\in\RR}$.
Moreover we derive the infinitesimal generator of $U_t$, i.e., we show that 
\eq{T}
\widehat U_t=e^{it\D(\dA)},\ \ \ t\in\RR,
\en 
where 
$$\D(\dA ):=\frac{i}{2}(\D^\dagger_T-\D_{\ov T})-i N_{\ov S},$$
and we prove that 
$\D(\dA)$ is essentially self-adjoint on a certain domain. 
\kak{T} also implies that 
$\widehat{U}_t$ gives the  normal-ordered quadratic expression of  $e^{it\dA}$.

In Berezin \cite[Chapter 3]{be} 
these  kinds of argument exist, however it is not rigorous at a few places. 
Quadratic operators such as $\D(A)$ has been studied in e.g., Araki \cite{ar1}, Araki-Shiraishi \cite{as} and  Araki-Yamagami \cite{ay}. 
Langmann \cite{la} calculated the local exponent 
 in \kak{q} in the different way as ours. 
 Fermionic cases for our discussion are  established by e.g., 
Lundberg \cite{lu}, Carey and Ruijsenaars \cite{caru} and Araki \cite{ar2}.

Next issue will be to study infinitesimal generators 
in the case where $T$ is {\it not} a Hilbert-Schmidt operator, i.e., 
\eq{S} A=\ST\not\in \sp_2.\en 
We, however, do not consider this problem here. 
Actually in the case where $T$ is not a Hilbert-Schmidt operator,  
we can not define $\dd T$ as an operator acting in $\fff$,  
and as was mentioned above the linear canonical transformation is not 
implemented by a unitary operator. 
So, in the case of \kak{S} we may have to shift our argument to white noise analysis. 
See \cite{kmp1,kmp2,kmp3} to this direction.

We organize this paper as follows. 
In Section 2 we review fundamental facts on the Fock space, quadratic operators.
In Section 3 we introduce one-parameter symplectic groups and main theorems. 
In Section 4 we show the weak differentiability of $U_t\Omega$ in $t$. 
In Section 5 we give proofs of the main theorems. 
In Section 6 we give some examples.

\section{Fundamental facts} 
\subsection{Boson Fock spaces}
Let $\fff=\fff(\hhh)$ denote the 
Boson Fock space over $\hhh$ defined  by 
$$\fff:=\bigoplus_{n=0}^\infty \fff^{(n)},$$ 
where $\fff^{(n)}=\hhh^{\otimes_s^n}$ is  the  $n$-fold symmetric tensor 
product of $\hhh$ with $\hhh^{\otimes_s^0} :=\BC$. 
Vector $\Psi$ of $\fff$ is written as 
$\Psi=\lkk \Psi^{(0)},\Psi^{(1)},\Psi^{(2)},\cdots\rkk$ 
with  $\Psi^{(n)}\in\fff^{(n)}$. 
The vacuum $\Omega\in\fff$ is defined by 
$$\Omega:=\{1,0,0,\cdots\}.$$
The creation operator 
$\add(f):\fff\rightarrow\fff$ 
smeared by $f\in\hhh$ is given  by 
$$\lk \add(f)\Psi\rk^{(n)}:=S_n (f\otimes\Psi^{(n-1)}),$$
where 
$S_n$ denotes  the symmetrizer of $n$-degree. 
Let 
$$\ffff:=
\mbox{the linear  hull of }\ 
\{\add(f_1)\cdots \add(f_n)\Omega|f_j\in\hhh,j=1,...,n,n\geq 0\}.$$
It is  known that $\ffff$ is dense in $\fff$. 
Simply for $f\in\hhh$, we write as $\ovv f$ for $\Gamma f$. 
 The annihilation operator $a(f)$ is defined by 
$$a(f):=\lk \left.\add(\ovv f)\right\lceil_\ffff\rk^\ast.$$
Since $\ass(f)\lceil_{\ffff}$ is closable, 
we denote its closed extension by the same symbol $\ass(f)$.
It holds that  
$$(\Psi,\add(f)\Phi)_\fff=(a(\ovv f)\Psi,\Phi)_\fff,\ \ \
 \Psi,\Phi\in\ffff.$$
where  $(f,g)_{\cal K}$ denotes the scalar product on Hilbert space 
${\cal K}$, 
which is linear in $g$ and antilinear in $f$. In addition, we  denote by
 $\|f\|_{\cal K}$ 
the associated norm. 
If no confusions arise, we omit ${\cal K}$ of $\|\cdot\|_{\cal K}$ 
and $(\cdot,\cdot)_{\cal K}$. 
The creation operator and the annihilation operator satisfy  
canonical commutation relations: 
\be
& & [a(f),\add(g)]=(\ovv f,g)_\hhh,\\
& & 
[a(f),a(g)]=0,\\
& & [\add(f),\add(g)]=0
\ee 
on $\ffff$. 
The field operator is  defined by 
$$\phi(f):=\frac{1}{\sqrt 2}(\add(\ovv f)+a(f)).$$      
The following proposition is known. 
\bp{facts}
(1) 
Suppose that a bounded operator $K$ commutes with $e^{i\phi(f)}$ for all
 $f\in\hhh$. 
Then 
$K$ is  a multiple of the identity.  

(2) Let ${\cal G}$ be a closed subspace of $\fff$ such that 
$e^{i\phi(f)}{\cal G}\subset {\cal G}$ for all $f\in\hhh$. 
Then  ${\cal G}=\fff$. 
\ep

\subsection{Symplectic groups}
Let $B=B(\hhh)$ denote the set of bounded operators on $\hhh$ and 
$\HS=\HS(\hhh)$  Hilbert-Schmidt operators. 
We denote the norm (resp. Hilbert-Schmidt norm) of a bounded operator 
$X$ on $\hhh$ by $\|X\|$ (resp. $\|X\|_2$).  
For  $S,T\in\B$ we define 
$$\A:=\ST:\hhh\oplus\hhh\rightarrow\hhh\oplus\hhh$$
by 
$$\A (\phi\oplus \psi):=(S\phi+\ov T \psi)\oplus (T\phi+\ov S \psi).$$
Let 
$$J:=\mat{\I}{0}{0}{-\I}:\hhh\oplus\hhh\rightarrow\hhh\oplus\hhh.$$
We define the   symplectic group 
$\sp$ and a subgroup $\bog$ of $\sp$ as follows. 
 \begin{definition}
(1) 
$$\sp:=\lkk \left. A=\ST:\hhh\oplus\hhh\rightarrow\hhh\oplus\hhh \right| 
\A J\A^\ast=\A^\ast J\A=J\rkk,$$
where 
$\A^\ast=
{\mat{S^\ast}{T^\ast}{\ov{T^\ast}}{\ov {S^\ast}}}.$

(2) 
$$\sp_2:=\lkk \left.A=\ST\in\sp\right|T \in B_2\rkk.$$
\end{definition}
Note that 
$\ov{(K^\ast)}=(\ov K)^\ast$ 
and that  the inverse  of  $\A\in\sp$ is given by 
\eq{25}
\A\f=J\A ^\ast J=\mat{S^\ast}{-T^\ast}{-\ov {T^\ast}}{\ov {S^\ast}}.
\en

\subsection{Quadratic operators} 
The number operator $N$ is defined by 
\be
&& D(N):=\lkk 
\left. \{\Psi^{(n)}\}\in\fff\right |\sum_{n=0}^\infty n^2\|\Psi^{(n)}\|^2_{\fff^{(n)}}<\infty\rkk,\\
&& (N\Psi)^{(n)}:=n\Psi^{(n)}.
\ee
Now we introduce  fundamental facts.  
\bp{funda}
(1) Let $f_1,...,f_m\in\hhh$. Then 
there exists a constant $c_m(f_1,...,f_m)$ such that 
for $\Psi\in D(N^{m/2})$, 
$$\|\add(f_1)\cdots \add(f_m)\Psi\|\leq c_m(f_1,...,f_m)  \|(N+1)^{m/2}\Psi\|.$$
Moreover suppose 
that $$f_j^{(n)}\rightarrow f_j,\ \ \ j=1,...,m,$$ strongly 
in $\hhh$ as $n\rightarrow \infty$ . Then  for $\Psi\in
 D(N^{m/2})$, 
\eq{l} s-\limn \ass(f_1^{(n)})...\ass(f_m^{(n)}) \Psi= 
\ass(f_1)...\ass(f_m) \Psi. 
\en
(2) Let $\{e_n\}_{n=1}^\infty$ be a complete orthonormal system.
Then 
\eq{ll}
(\Psi, N\Phi)=\sum_{n=1}^\infty(a(e_n) \Psi, a(e_n)\Phi).
\en 
\ep
\proof See e.g., \cite[Section X.7]{rs2}. 
\qed
Let $K\in\HS$. Then there exist two orthonormal systems 
$\{\psi_n\}$, $\{\phi_n\}$ in $\hhh$,  and a positive sequence 
$\lambda_1\geq \lambda_2\geq...>0$ such that 
$$Kf=\sumn \la_n (\psi_n,f)\phi_n,\ \ \ f\in\hhh,$$
with 
$\sumn\la_n^2=\|K\|^2_2.$
\bl{funda2}
Let $K\in B_2$ and $S\in B$. 
Then for $\Psi\in\ffff$, 
\bee
&&(1) \| \sumnn\la_n\add(\ovv\psi_n)\add(\phi_n)\Psi\|\leq \sqrt 6\|K\|_2\|(N+1)\Psi\|\\
&&(2) \| \sumnn\la_n a(\ovv\psi_n) a(\phi_n)\Psi\|\leq \|K\|_2\|N\Psi\|,\\
&&(3) \| \sumnn \add(e_n)a(\ovv{S^\ast e_n})\Psi\|\leq \|S\|\|N\Psi\|, 
\eee
where $\{e_n\}$ is a complete orthonormal system of $\hhh$. 
\el
\proof 
We have 
\be
 \|\sumnn\la_n\add(\bsn)\add(\fn) \Psi\|^2
&=&\sum_{n,m}^M \lambda_n\lambda_m 
(\add(\bsm)\add(\fm)\Psi, \add(\bsn)\add(\fn)\Psi)\\
&=&A+B+C+D+E,
\ee 
where 
\be 
&&A=  \sum_{n}^M \lambda_n^2
(\Psi, a(\bsn)\add(\sn)\Psi),\\
&&B=\sum_{n}^M \lambda_n^2 
(\Psi, \add(\bfn)a(\fn)\Psi),\\
&&C=\sum_{n}^M \lambda_n (\Psi, a(\ov K \sn)\add(\bfn)\Psi),\\
&&D=\sum_{n}^M \lambda_n (\Psi, \add(\sn)a(K^\ast\bfn)\Psi),\\ 
&&E=\sum_{n}^M \lambda_n (a(\bsn) a(\fn)\Psi, a(\bsm) a(\fm) \Psi).
\ee
We have 
$$
|A|\leq\sum_n^M\lambda_n^2\|\add(\sn)\Psi\|^2
\leq \sum_n^M\lambda_n^2\|(N+1)^\han \Psi\|^2
\leq 
\|K\|_2^2 \|(N+1)^\han \Psi\|^2, 
$$
and 
$$
|B|\leq 
\sum_n^M\lambda_n^2\|a(\fn)\Psi\|^2
\leq 
\sum_n^M\lambda_n^2\|N^\han \Psi\|^2
\leq
\|K\|_2^2 \|N^\han \Psi\|^2.
$$
We have 
\be 
C&=& 
\sum_n^M\lambda_n (K\bsn,\bfn)\|\Psi\|^2+
\sum_n^M\lambda_n(\Psi, \add(\bfn)a(\ov K \sn)\Psi)
\ee
We estimate the right-hand side above such as 
\be
\left| 
\sum_n^M\lambda_n (K\bsn,\bfn)\|\Psi\|^2\right|
&\leq &
\sum_n^M\lambda_n \|K\bsn\|\|\Psi\|^2\\
&\leq & \lk \sum_n^M\lambda_n^2 \rk^\han \lk\sum_n^M \|K\bsn\|^2\rk^\han 
\|\Psi\|^2\\
&\leq & \|K\|_2^2\|\Psi\|^2,
\ee
and 
\be 
|\sum_n^M\lambda_n(\Psi, \add(\bfn)a(\ov K \sn)\Psi)|
&\leq&  \lk \sum_n^M\lambda_n^2\rk^\han \lk \sum_n^M 
\|a(\fn)\Psi\|^2 \|K\|^2\|N^\han\Psi\|^2\rk^\han\\
&\leq &\|K\|_2\|K\|\|N^\han\Psi\|^2.
\ee
Hence 
$$|C|\leq\|K\|_2^2\|\Psi\|^2+\|K\|_2\|K\|\|N^\han\Psi\|^2.$$ 
We have 
\be
|D|&\leq & \lk\sum_n^M\lambda_n^2\rk^\han \lk \sum_n^M\|a(\bsn)\Psi\|^2\|K\|^2\|N^\han\Psi\|^2\rk^\han\\
&\leq & \|K\|_2\|K\|\|N^\han\Psi\|^2.
\ee
Finally  we see that 
\be 
|E|&=&\|\sum_{n=1}^M\lambda_na(\bfn)a(\sn)\Psi\|^2\\
&\leq &\sum_{n=1}^M\lambda_n^2\sum_{n=1}^M\|a(\bfn)a(\sn)\Psi\|^2\\
&\leq & \|K\|_2^2 \sum_{n=1}^M\|N^\han a(\sn)\Psi\|^2\\
&= & \|K\|_2^2 \sum_{n=1}^M (a(\sn)\Psi, Na(\sn)\Psi) \\
&= & \|K\|_2^2 \sum_{n=1}^M (a(\sn)\Psi, a(\sn) N \Psi)-
(a(\sn)\Psi, a(\sn)\Psi) \\
&\leq & \|K\|_2^2  (\Psi,  N^2 \Psi)\\
&=& \|K\|_2^2 \|N\Psi\|^2.
\ee 
Combining estimates from  $A$ to $E$ above, we obtain that 
\be 
&&\| \sumnn\la_n\add(\ovv\psi_n)\add(\phi_n)\Psi\|^2\\
&&\leq \|K\|_2^2\lk\|(N+1)^\han\Psi\|^2+3\|N^\han\Psi\|^2+\|\Psi\|^2+\|N\Psi\|^2\rk\\
&&\leq 6\|K\|_2^2\|(N+1)\Psi\|^2.
\ee
Thus (1) follows.
From the estimate of  E in (1) we have 
\be
\|\sumnn\la_n a(\bfn)a (\sn)\Psi\|^2&\leq&\|K\|_2^2\|N\Psi\|^2.
\ee
Thus (2) follows.
Finally we estimate (3). 
It is  proven on $\ffff$ that 
$$s-\lim_{M\rightarrow \infty}\sumnn\add(e_n)a(\Gamma S^\ast e_n)\Psi=
\bigoplus_{n=1}^\infty \left[
\sum_{j=1}^n 1\otimes\cdots\otimes \stackrel{j}{S}\otimes\cdots\otimes 1\right]\Psi^{(n)}.$$
Since 
$$\|\bigoplus_{n=1}^\infty \left[
\sum_{j=1}^n 1\otimes\cdots\otimes \stackrel{j}{S}\otimes\cdots\otimes 1\right]\Psi^{(n)}\|
\leq 
\|S\|\|N\Psi\|,$$
we obtain (3). 
\qed
From Lemma  \ref{funda2} 
we can define for $\Psi\in\ffff$ 
\bee 
\eqq{ss1}
&&\dd K\Psi:=s-\limM\sumnn\la_n\add(\ovv\psi_n)\add(\phi_n)\Psi,\\
&&
\eqq{ss2}
\aa K\Psi:=s-\limM\sumnn\la_n a(\ovv\psi_n) a(\phi_n)\Psi,\\
&&
\eqq{ss3}
\da S\Psi:=s-\limM\sumnn \add(e_n)a(\ovv{S^\ast e_n})\Psi. 
\eee
Let $\Psi=\add(f_1)\cdots\add(f_n)\Omega$. 
Then it is seen that 
\bee
\label{1}
& & \aa K \Psi= \sum_{i\not=j}^n(\bar f_i, (K+\T K) f_j)  
\add(f_1)\cdots \widehat{\add}(f_i)\cdots \widehat{\add}(f_j)\cdots  \add (f_n)\Omega,\\
\label{2}
& & \da S \Psi=\sum_{j=1}^n 
\add(f_1)\cdots{\add}(Sf_j)\cdots  \add (f_n)\Omega. 
\eee
The following lemma  is inherited from Lemma \ref{funda2}. 
\bl{funda3}
Let $K\in B_2$ and $S\in B$. Then for $\Psi\in\ffff$, 
\bee
\label{ss11}
&& \|\dd K \Psi\|\leq \sqrt6\|K\|_2\|(N+1)\Psi\|,\\
\label{ss12}
&& \|\aa K\Psi\|\leq  \|K\|_2\|N \Psi\|,\\
\label{ss13}
&& \|\da S\Psi\|\leq \|S\|\|N\Psi\|.
\eee
\el
 From this lemma it is shown that the domains of $\dd K$, $\aa K$ and $\da S$ can be extended to $D(N)$.  
We directly see the following lemma. 
\bl{san}
\begin{itemize}
\item[(1)] We have 
$(\D_K)^\ast=\D_{K^\ast}^\dagger$ and 
$(N_S)^\ast=N_{S^\ast}$ on $D(N)$.
\item[(2)] 
\bee
\label{u1}
& & [\dd K, a(f)]=-\add((K+\T K) f),\\
& & 
\label{u2}
[\aa{K}, \add(f)]=a((K+\T K) f),\\
& & 
\label{u3}
[\da S, a(f)]=-a(\ov{S^\ast} f),\\
& & 
\label{u4}
[\da S,\add(f)]=\add( S f).
\eee
\item[(3)]
Let $K,L\in B_2$ and $S, T\in B$. Then on $D(N)$,  
\be 
&& \dd{K+L}=\dd K+\dd L,\\
&& \aa {K+L}=\aa K+\aa L,\\
&& \da {S+T}=\da S+\da T.
\ee
\ei
\el
From above commutation relations it follows that 
\eq{maru1}
\|\dd K\Omega\|^2= {\rm Tr}[K^\ast(K+\T K)].
\en 
We set 
$$\ddd:=\bigcap_{k=1}^\infty D(N^k).$$
\bl{11}
(1) Suppose that 
$$(i)\ K\in\HS,\ \ \ 
(ii)\ \ov {K^\ast}=K,\ \ \ 
(iii)\ \|K\|<1.$$
Then for $\Psi\in\ffff$, 
$$\ui{K}\Psi:=s-\limM\sumnnm\frac{1}{n!}\lk-\half \dd K\rk^n\Psi$$
exists and 
$\ui{K}\Psi\in\ddd$. \\
(2) 
Suppose that $S\in \B$ and $K\in\HS$. 
Then for $\Psi\in\ffff$, 
\bee
& & {\uii{S}}\Psi:=s-\limM\sumnnm\frac{1}{n!}\wick{\lk-\half \da S\rk^n} \Psi
\\
& & 
{\uiii{K}}\Psi:=s-\limM\sumnnm\frac{1}{n!}\lk-\half \aa K\rk^n\Psi
\eee 
exist, and 
${\uii{S}}\Psi\in\ffff$ and $\uiii{K}\Psi\in \ffff$. 
\el
\proof 
We shall prove (1). 
It is enough to show the lemma for $\Psi=\add(f_1)\cdots \add(f_N)\Omega$. 
Let 
$$a_n=\lk \frac{1}{2^n}\frac{1}{n!}\rk^2 \|({\dd K})^n\Omega\|^2.$$
It is proven in \cite{rui} that 
$f(z)=\sum_{m=0}^\infty a_m z^m$ exists for $|z|<1/\|K\|^2$ 
and 
$$f(z)={\rm det}(1-zK^\ast K)^{-\han}.$$ 
In particular $f(z)$ is analytic for $|z|<1/\|K\|^2$. 
We have 
\be \|\sum_{m=0}^\infty\frac{1}{m!} (-\half\dd K)^m \Psi\|^2 
&\leq & 
\|f_1\|^2\cdots\|f_N\|^2\\
&&\times \sum_{m=0}^\infty (N-1+2m)(N-2+2m)\cdots 2m a_m<\infty.
\ee
Thus (1) is proven. 
(2) follows from the  fact that 
$\wick{\da S^n}\Psi=0$ and $\aa K^n\Psi=0$ for a sufficiently large $n$. 
\qed
From \kak{u1}-\kak{u4} the following  commutation relations hold on $\ffff$: 
\bee
\label{tt1}
& & [\ui K, a(f)]=\half \add((K+\T K) f) \ui K,\\
& & 
\label{tt4}
[\uiii K, \add(f)]=- \half a((K+\T K) f)\uiii K,\\
& & 
\label{tt2}
[\uii S, a(f)]=\half a(\ov{S^\ast} f)\uii S,\\
& & 
\label{tt3}
[\uii S, \add(f)]=-\half \add(S f)\uii S.
\eee

\subsection{Proper canonical  transformations}
Let $\A=\ST$. Then $\A$  induces  the following maps:  
\bee
\label{a1}
& & a(f)\longmapsto \bbb(f):=a(Sf)+\add(Tf),\\
\label{a2}
& &  \add(f)\longmapsto \bdd(f):=a(\ov Tf)+\add(\ov Sf).
\eee
Formally we may write 
$$(
{\bbb (f)} \ \ 
{\bdd(f)})
=({a(f)}\ \ {\add(f)})\ST.$$ 
Then canonical commutation relations 
\bee
\label{r1}
& & [\bbb(f),\bdd(g)]=(\ovv f, g)_\hhh,\\
\label{r2}
& & [\bbb(f),\bbb(g)]=0,\\
\label{R3}
& & 
[\bdd(f),\bdd(g)]=0,
\eee 
hold on $\ffff$ and 
$$(\Psi, \bdd(f)\Phi)_\fff =(\bbb(\ovv f)\Psi,\Phi)_\fff ,\ \ \ 
\Psi,\Phi\in\ffff,$$
follows. 
Since $\bss(f)\lceil_{\ffff}$ is closable, we denote its closed extension
 by the same symbol $\bss(f)$. 
The following proposition  is well known. 
\begin{proposition}
There exists a unitary operator $\UB(A)$ on $\fff$ 
such that 
$$\UB(A):D(\ass(f))\rightarrow D(\bss(f))$$ with 
\eq{ua}
\UB(A)\ass(f)\UB(A)\f=\bss(f)
\en 
if and only if 
$\A\in\bog$. 
\end{proposition}
\proof See \cite{be,rui}.
\qed
If $\UB'(A)$ also implements such as \kak{ua}, 
then 
it holds that 
$$\UB(A)\f\UB'(A)\ass(f)=\ass(f)\UB(A)\f\UB'(A).$$
Hence 
$$\UB(A)\f\UB'(A)e^{i\phi(f)} =e^{i\phi(f)} \UB(A)\f\UB'(A),$$
which implies that 
$$\UB'(A)=\omega \UB(A)$$
with some $\omega\in\{e^{i\psi}| \psi\in\RR\}$ 
by Proposition \ref{facts}.

Now we  construct a  unitary operator $\UB(A)$ 
concretely. 
The condition $A=\ST\in \sp$  is equivalent with the following algebraic relations:  
\bee
\eqq{s1}
& & S^\ast S-T^\ast T=\I,
\\
& & 
\eqq{s2}\ov {S^\ast} T-\ov {T^\ast} S=0,
\\
& & \eqq{s3}SS^\ast-\ov {T {T^\ast}}=\I,
\\
& & \eqq{s4}TS^\ast-\ov{ S  T^\ast} =0.
\eee
\bl{alg}
Let $\A=\ST\in\sp$. Then 
(1)  $S\f\in B$, 
(2)  $\|TS\f\|<1$, 
(3)  $\ov{(TS\f)^\ast}=TS\f$, 
(4)  $\ov{(S\f \ov T)^\ast} =S\f \ov T$. 
\el
\proof
From \kak{s1} it follows that 
\eq{s5}
S^\ast S=\I+T^\ast T\geq \I.
\en
Thus $\|S\|^2\geq 1$, and (1) follows.
 By \kak{s5} we have 
$TS\f=T(\I+T^\ast T)\f S^\ast,$
which implies that
$$ 
(TS\f)(TS\f)^\ast=
T(\I+T^\ast T)\f S^\ast S(\I+T^\ast T)\f T^\ast=T(\I+T^\ast T)\f T^\ast.$$ 
Thus 
\be\|TS\f\|^2 &=&   \|T (\I+T^\ast T)\f T^\ast \|\\
&=& \| T(1+T^\ast T)^{-1/2} (1+T^\ast T)^{-1/2} T^\ast\|\\
&=&
\| (1+T^\ast T)^{-1/2} T^\ast T (1+T^\ast T)^{-1/2} \|<1.
\ee
Thus (2) follows. 
By \kak{s2} we have 
$\oa S T S\f=\oa T.$ 
Then $S^\ast \ov{TS\f}=T^\ast$  follows. 
Note that $(S^\ast)\f=(S\f)^\ast$. It is obtained that 
$$TS\f=\ov{(S^\ast)\f T^\ast}=\ov{(S\f)^\ast T^\ast}=\ov{(TS\f)^\ast}.$$
Hence (3) follows. 
Similarly (4) is obtained  from \kak{s4}.
\qed
Let 
$\A=\ST\in\bog$.
We set 
$$K_1:=TS\f,\ \ \ K_2:=\I-\ov{(S\f)^\ast},\ \ \ 
K_3:=-S\f \ov T.$$
Since $K_1\in \HS$, $\ov {K_1^\ast}=K_1$  and $\|K_1\|<1$, 
we see that by Lemma \ref{11},  
\eq{for}
\U(\A):={\rm det}(\I-K_1^\ast K_1)^{1/4}
\ui{K_1}
{\uii{K_2}} 
\uiii{K_3}
\en 
is  well defined on $\ffff$. 
Moreover it is seen that 
$$\U(A)\Psi\in {\cal D}_\infty,\ \ \ \Psi\in\ffff.$$
\bl{bog}
Let $\A\in\bog$. Then 
$\U(\A)$ can be uniquely extended to a unitary operator on $\fff$. 
\el
\proof  
By the commutation relations \kak{tt1}-\kak{tt4} it is seen that 
\eq{OY}
U(A) \ass(f)U(A)\f \Psi= \bss(f)  \Psi
\en  for 
$\Psi\in\ffff$.
From this,  and the canonical commutation relations \kak{r1} and \kak{r2}, 
 it follows that 
 \be
&&\|\U(\A)\add(f_1)\cdots \add(f_n)\Omega\|^2\\
&&=
\|\bdd(f_1)\cdots \bdd(f_n)\U(\A)\Omega\|^2\\
&&=
{\rm det}(1-K_1^\ast K_1)^{\han} 
\|\bdd(f_1)\cdots \bdd(f_n)\ui{K_1}\Omega\|^2\\
&&=
{\rm det}(1-K_1^\ast K_1)^{\han} \sum_{\pi\in{\cal P}_n}(\ovv  f_1, f_{\pi(1)})\cdots
(\ovv  f_n, f_{\pi(n)}) \|\ui{K_1}\Omega\|^2\\
&&=
\|\add(f_1)\cdots \add(f_n)\Omega\|^2,
\ee
where ${\cal P}_n$ denotes the set of permutations of $n$ degree, 
and  we used that 
$$\|\ui{K_1}\Omega\|^2={\rm det}(1-K_1^\ast K_1)^{-\han}$$
and 
$$b_A(f) e^{-\half \D_{K_1}^\dagger}\Omega=0,\ \ \ f\in\hhh.$$
Then $\U(\A)$ is an isometry from $\ffff$ onto ${\cal E}$, where 
\be
&& 
{\cal E}:= \mbox{the linear hull of }\times\\
&&\hspace{2cm}\times
\{\bdd(f_1)\cdots \bdd(f_n)\U(\A)\Omega,\U(A)\Omega| f_j\in\hhh, j=1,...,n,n\geq 1\}.
\ee 
From \kak{25}  
it follows that 
\eq{inverse}
(a(f)\ \ {\add(f)})
=(\bbb(f)\ \ \bdd(f))
\mat{S^\ast}{-T^\ast}{-\ov {T^\ast}}{\ov {S^\ast}}.
\en 
By this  we see that 
\eq{da1}
\ass(f) {\cal E}\subset {\cal E},\ \ \ f\in\hhh.
\en 
Let $\Psi\in{\cal E}$ and 
$$\Psi_N:=\{\Psi^{(0)},\Psi^{(1)}, \Psi^{(2)}, ...,\Psi^{(N)},0,0,...\}.$$
Since $\Psi\in\ffff$, we see that 
$\Psi_N$ is an analytic vector of $\phi(f)$, i.e., 
\eq{da2}
e^{i\phi(f)}\Psi_N=\sumn \frac{1}{n!}(i\phi(f))^n\Psi_N,
\en 
which implies, together with \kak{da1}, that 
$e^{i\phi(f)}\Psi_N\in\ov{\cal E}$, and by a limiting argument 
$e^{i\phi(f)}\Psi\in\ov{\cal E}$.
Thus 
$e^{i\phi(f)}{\cal E}\subset \ov{\cal E}$ follows. 
By a limiting argument we have 
$$e^{i\phi(f)}\ov {\cal E}\subset \ov{\cal E}.$$
Thus 
$\ov{\cal E}=\fff$ by Proposition \ref{facts}. 
Hence we conclude that 
$\U(\A)$ can be uniquely extended to a unitary operator on $\fff$. 
The lemma follows. 
\qed
We denote the  unitary extension of $\U(A)$ by the same symbol $\U(\A)$. 
\bp{228}
Let $A\in\bog$. Then 
we can choose  $\U(A)$ as $\UB(A)$ in \kak{ua}. 
\ep
\proof
Note that 
$D(\ass(f))\supset {\cal D}_\infty\supset\ffff.$
In particular ${\cal D}_\infty$ is a core of $\bss(f)$. 
Since $\U(A):\ffff\rightarrow {\cal D}_\infty$, 
we see that $\U(A)$ maps a core of $\ass(f)$ to a core of $\bss(f)$. 
By \kak{OY} we conclude that $\U(A)$ maps $D(\ass(f))$ to $D(\bss(f))$ with
$$\ass(f)=\U(A)\f \bss(f) \U(A). $$ 
Thus the proposition follows.
\qed
$U(\cdot)$ gives a projective unitary representation of $\bog$, i.e., 
$$U(A)U(B)=\omega(A,B)U(AB)$$
with a  local factor $\omega(A,B)\in\{e^{i\psi}|\psi\in\RR\}$.

\section{One-parameter unitary groups}
\subsection{One-parameter symplectic groups}

\begin{definition}
\label{H}
\be
\SS:
&=& \lkk \dA=\ST\left| 
\dA J+J\dA=\dA^\ast J+J\dA^\ast=0, T\in B_2\right.\rkk\\
&=& 
\lkk \dA=\ST\left| 
S^\ast=-S, \ov{T^\ast}=T,   T\in B_2
\right.\rkk.
\ee
\end{definition}

\bl{Fumio}
Let $\dA \in\SS $. 
Then 
$$e^{t\dA }\in\bog,\ \ \ t\in\RR.$$
\el
In order to prove  Lemma \ref{Fumio} we need  lemmas.

\bl{one}
We have  $e^{t\dA }\subset \sp$ if and only if $\dA J+J\dA=\dA^\ast J+J\dA^\ast=0$.
\el
\proof
Assume that $e^{t\dA }\in \sp$. Namely 
\eq{t}
e^{t\dA }Je^{tA^\ast}=e^{tA^\ast} J e^{t\dA } =J,\ \  \ t\in\RR.
\en 
Take a strong derivative at $t=0$ on the both sides of \kak{t}. 
Then 
it follows that 
\eq{A}
AJ+JA^\ast=A^\ast J+JA=O. 
\en 
Conversely assume \kak{A}. 
Then we have 
$$\frac{d}{dt} e^{t\dA }Je^{tA^\ast}=0.$$
Hence it holds that 
$$J(t):=e^{t\dA }Je^{tA^\ast}=J(0)=J,\ \ \ t\in\RR .$$
In the similar manner 
$$e^{tA^\ast}Je^{t\dA }=J,\ \ \ t\in\RR $$
is proven. Thus the lemma follows. 
\qed
For $\dA\in\SS $, 
we set 
$$e^{t\dA }=\mat
 {(e^{t\dA })_{11}}{(e^{t\dA })_{12}}{(e^{t\dA })_{21}}{(e^{t\dA })_{22}}
=
\mat {(e^{t\dA })_{11}}{\ov {(e^{t\dA })_{21}}}{(e^{t\dA })_{21}}{\ov{(e^{t\dA })_{11}}}.$$
In what follows we set 
$\Sdd=\Sd$ and $\Tdd=\Td$. 
Moreover $\BT$ denotes the set of Hilbert-Schmidt operators 
on $\hhhh$, and 
$\|X\|_\BT $ 
 the Hilbert-Schmidt norm of $X\in\BT$.

\bl{31}
Let $\dA =\ST\in\SS $. 
Then 
$$e^{t\dA }-e^{t\Sdd}\in \BT,\ \ \ t\in\RR,$$ 
and 
$$\left\| 
e^{t\dA }-e^{t\Sdd} 
\right\|_{\BT}  \leq \frac{\|{\bf T} \|_{\BT} }{\|T\|} \lk 
e^{t \|T\|}-1\rk,\ \ \ t\in\RR.$$
\el
\proof
Let 
\eq{39}
Y(t)=e^{t\dA } e^{-t\Sdd},\ \ \ 
T(t)=e^{t\Sdd} \Tdd e^{-t\Sdd}=
\mat 0 {e^{tS}\ov T e^{-t\ov S}} {e^{t\ov S } T e^{-tS}} 0.
\en 
We see that 
$$\frac{d}{dt} Y(t)=Y(t)T(t)$$
in the operator norm on $\hhhh$, 
which implies that 
\eq{ynt}
Y(t)=E+\sum_{n=1}^\infty Y_n(t),
\en 
where 
$$ 
Y_n(t)=\intt T(t_1)T(t_2)\cdots T(t_n) \dttt.
$$ 
We have 
$$\hspace{-2cm}
\|Y_n(t)\|_\BT\leq 
\intt \|T(t_1)\|_{\BT}\times $$
$$\hspace{2cm}\times 
\|T(t_2)\|_\hhhh \cdots \|T(t_n)\|_\hhhh  \dttt,$$
where $\|X\|_\hhhh$ denotes the norm of bounded operator $X$ on $\hhhh$. 
Since 
$e^{t\Sdd}$ is a unitary operator in $\hhh\oplus\hhh$, 
we have 
\be
&& \|T(t_j)\|_\hhhh \leq  \|T\|,\ \ \, j=2,3,...,n, \\
&&
\|T(t_1)\|_{\BT}\leq  \|\bT\|_{\BT} .
\ee
Then it follows that
\eq{ynt2}
\|Y_n(t)\|_{\BT}\leq \frac{\|T\|^{n-1}}{n!} t^n  \|\bT\|_{\BT} ,
\en 
which yields that 
$Y(t)-E\in \BT$
and 
\eq{F1}
\|Y(t)-E\|_{\BT}\leq \frac{\|\bT\|_{\BT} }{\|T\|}\lk e^{t\|T\|}-1\rk.
\en
From \kak{F1} it follows that 
\be \left\| e^{t\dA }-e^{t\Sdd}\right\|_{\BT}
&=& \|(Y(t)-E)e^{t\Sdd}\|_{\BT} \\
&\leq& \|(Y(t)-E)\| _{\BT} \\
&\leq& 
\frac{\|\bT\|_{\BT} }{\|T\|}\lk e^{t\|T\|}-1\rk.
\ee 
Hence the lemma follows.
\qed

{\it Proof of Lemma \ref{Fumio}}\\
The fact 
$e^{t\dA }\in\sp$ follows from Lemma \ref{one}. 
We shall prove 
$(e^{t\dA })_{21}\in B_2$. 
By Lemma \ref{31}, 
$e^{t\dA }-e^{t\Sdd}\in \BT.$
It implies 
$$(e^{t\dA }-e^{t\Sdd})_{21}\in B_2.$$
Since 
$(e^{t\dA })_{21}=(e^{t\dA }-e^{t\Sdd})_{21}$, 
$e^{t\dA}\in\bog$ follows.
\qed

\begin{remark}
In \cite[Lemma 6.3]{be} it has been proven that 
if 
$S, T\in B$, $S^\ast=-S$,  $\ov{T^\ast}=T$, and 
$$
F(t)=\int_0^t e^{\tau \ov{S}} T e^{-\tau S}d\tau \in B_2
$$ with 
$ 
\| F(t)\|_2\in L_{\rm loc}^1(\RR,dt).$ 
Then 
$e^{t\dA }\in\bog$.
\end{remark}

\subsection{The main theorems}
For $A\in\SS $, we set 
$$U_t:=U(e^{t\dA}),\ \ \ t\in\RR.$$
$U_t$ gives a projective unitary representation of 
$\{e^{t\dA}\}_{t\in\RR}$, i.e., 
$$U_tU_s=e^{i\rho(t,s)}U_{t+s},$$
where 
we set 
$$e^{i\rho(t,s)}:=\omega(e^{t\dA}, e^{s\dA}).$$
For $\dA=\ST$, let 
$$\D(\dA ) :=\frac{i}{2}( \D_T^\ast-\D_{\ov T})-iN_{\ov S}.$$
The main theorems in this paper are as follows. 
\bt{mmm}
 Let $\dA \in \SS $. 
Then 
$\D(\dA )$ is essentially self-adjoint on $\ffff$. 
\et
\bt{mm}
 Let $\dA \in \SS $. 
Then 
$$e^{it\D(\dA )} \ass(f) e^{-it\D(\dA )} =b^\sharp(f),\ \ \ t\in\RR.$$ 
\et
For notational convenience we set 
\be
\K:=\TT \OO\f.
\ee
\bt{mM}
Let $\dA \in\SS $. Then 
$$
\U_t=e^{i\int_0^t \TA(s) ds} e^{it\D(\dA )},\ \ \ t\in\RR,$$
where 
$$\TA(s):=\half\Im{\rm Tr}(T^\ast K_s).$$
\et

\bc{PP}
Let $\dA \in\SS $. $\int_0^t \Im {\rm Tr}(T^\ast K_s)ds =0$ if and only if 
$$\U_t=e^{it\D(\dA )},\ \ \ t\in\RR.$$
Namely in the case of $\int_0^t \Im {\rm Tr}(T^\ast K_s)ds=0$, 
$U_t$ gives a unitary representation of $\{e^{t\dA}\}_{t\in\RR}$. 
\ec
\proof 
It follows from Theorem \ref{mM}.
\qed

\bc{PPPP}
Let $\dA\in\SS$. Then the local exponent $\rho(\cdot,\cdot)$   is  given by 
$$\rho(t,s)= \int_0^t\TA(r)dr+\int_0^s\TA(r) dr-\int_0^{t+s}\TA(r) dr.$$ 
\ec
\proof 
We directly see that 
$$U_tU_s=e^{i\lkk \int_0^t\TA(r)dr+\int_0^s\TA(r)dr\rkk}e^{i(t+s)\D(\dA)}=
e^{i\lkk \int_0^t\TA(r)dr+\int_0^s\TA(r)dr-\int_0^{t+s}\TA(r)dr\rkk}U_{t+s}.$$
Then the corollary follows.
\qed

\bc{PPP}
Let $\dA\in\SS$. Then 
$$\widehat U_t:=e^{-i\int_0^t \TA(s) ds} U_t$$ 
gives a unitary representation of $\{e^{t\dA}\}_{t\in\RR}$, 
 and $\widehat U_t$ is 
a normal ordered quadratic expression of $e^{it\D(A)}$. In particular 
it follows that 
$$(\Omega, e^{it\D(A)}\Omega)=
{\rm det}(1-K_t^\ast K_t)^{1/4} e^{-i\int_0^t\TA(s)ds}.$$
\ec
\proof 
Since we actually have 
$\widehat U_t=e^{it\D(A)}$, the corollary follows.
\qed

\begin{example}
Let  
\eq{DS}
\ov S=S=-S^\ast,\ \ \ \ov T=T={T^\ast}.
\en 
Then  $\dA=\ST$ satisfies the assumptions  of Corollary \ref{PP}.

Let $\hhh=L^2(\RR,dx)$ and $\Gamma$ be the complex conjugation. 
Define the Hilbert-Schmidt operator $T$ by 
$$T\phi(x)=\int_\RR g(x,y)\phi(y) dy$$ 
with 
a real-valued function $g(x,y)\in L^2(\RR\times\RR)$. 
Let $f$ be  a real-valued measurable  function such that 
$f\in  L^\infty(\RR)$ and $f(-x)=-f(x)$. 
Define  $S$ by 
$$ S= f(\frac{d}{dx} ).$$
Then $S$ and $T$ satisfy  \kak{DS}.
\end{example}

\section{Weak differentiability  of $U_t$} 
In this section we shall prove 
the weak differentiability of $U_t\Omega$ 
in $t$.
Throughout this section we assume that 
$\dA=\ST\in\SS$. 
The next lemma is fundamental. 
\bl{L0}
Let $T_t\in B_2$ and $S_t\in B$. 
Assume that $T_t$ is differentiable in $t$ in $B_2$, and $S_t$ in $B$ with 
$\displaystyle \frac{d}{dt}T_t=T_t'$ and $\displaystyle \frac{d}{dt}S_t=S_t'$. 
Then 
$T_t S_t$ is differentiable in $t$ in $B_2$ with 
$$\frac{d}{dt} T_t S_t=T_t' S_t+T_t S_t'.$$
\el
\proof 
We have 
\be
&& \left \|
\frac {T_{t+\epsilon}S_{t+\epsilon}-T_tS_t}{\epsilon}-T_t'S_t-T_tS_t'\right\|_2\\
&& \leq \left 
\|\frac{T_{t+\epsilon}-T_t}{\epsilon}-T_t'\right\|_2
\|S_{t+\epsilon}\|+\|T_t'\|_2\|S_{t+\epsilon}-S_t||
+\|T_t\|_2
\left\|\frac{S_{t+\epsilon}-S_t}{\epsilon}-S_t'\right\|\\
&&\rightarrow 0
\ee
as $\epsilon\rightarrow 0$. Hence the lemma follows.  
\qed
\bl{L1}
It follows that 
$(e^{t\dA})_{ij}$, $i\not=j$ (resp. $i=j$), is differentiable in $t$ in $B_2$ (resp. $B$) 
with 
$$\frac{d}{dt} (e^{t\dA})_{ij}=(\dA e^{t\dA})_{ij}.$$ 
\el
\proof
Since 
$$\frac{d}{dt} e^{t\dA}=\dA e^{t\dA}\ \ \ 
in\ B(\hhh\oplus\hhh),$$ 
it follows that 
$$\frac{d}{dt}(e^{t\dA})_{ij}=(\dA e^{t\dA})_{ij}\ \ \ 
in\  B.$$ 
Let $i=2,j=1$. 
Then we have 
\be
&& \left
\|\lkk \frac{1}{\epsilon}\lk e^{(t+\epsilon)\dA}-e^{t\dA}\rk-\dA e^{t\dA}\rkk_{21}\right\|_2\\
&&=
 \left\| \sum_{k=1,2} 
(e^{t\dA})_{2k} \lkk 
 \frac{1}{\epsilon}\lk e^{\epsilon\dA}-E\rk-A\rkk_{k1}\right\|_2\\
&&\leq 
 \| (e^{t\dA})_{21}\|_2 
\left\| \lkk 
 \frac{1}{\epsilon}\lk e^{\epsilon\dA}-E\rk-A\rkk_{11}\right\|+
 \| (e^{t\dA})_{22}\| 
\left\| \lkk  \frac{1}{\epsilon}\lk e^{\epsilon\dA}-E\rk-A\rkk_{21}\right\|_2.
\ee
Since 
\be 
 \left \| \lkk \frac{1}{\epsilon}\lk e^{\epsilon \dA}-E\rk-A\rkk_{11}\right \|\rightarrow0,\ \ \ 
\|(e^{\epsilon \dA})_{22}\|\rightarrow 1
\ee
as $\epsilon\rightarrow 0$, it is enough to prove that 
\be
&& 
\left\| \lkk  \frac{1}{\epsilon}\lk e^{\epsilon\dA}-E\rk-A\rkk_{21}\right\|_2=
\left\|   \frac{1}{\epsilon} (e^{\epsilon\dA})_{21}-T\right\|_2\rightarrow 0
\ee
as $\epsilon\rightarrow 0$.
By \kak{39} and \kak{ynt}  
we see that 
$$\mat{0}{\ov{(e^{t\dA })_{21}}} 
{{(e^{t\dA })_{21}}} {0}
=\sum_{m=0}^\infty Y_{2m+1}(t) e^{t{\bf S}} .$$
Hence 
$$\mat{0}{\ov{(e^{t\dA })_{21}}-t\ov T} 
{{(e^{t\dA })_{21}}-t T} {0}
=\sum_{m=0}^\infty Y_{2m+1}(t) e^{t{\bf S}}-t{\bf T} .$$
Since 
$$ \ll  \mat{0}{\ov{(e^{t\dA })_{21}}-t\ov T} 
{{(e^{t\dA })_{21}}-t T} {0}\rr _\BT ^2 =2\|(e^{t\dA })_{21}-tT\|_2^2,$$
it is enough to show that 
\eq{SH}
\limt \frac{1}{|t|} \ll 
\sum_{m=0}^\infty Y_{2m+1}(t) e^{t{\bf S}}-t{\bf T}
\rr_\BT =0.
\en 
Since $\|e^{t{\bf S}}-E\|_{\hhhh}= \|e^{tS}-1\|$, we have 
$$\ll \sum_{m=0}^\infty Y_{2m+1}(t)e^{t{\bf S}}-t{\bf T}\rr _\BT\leq 
\ll \sum_{m=0}^\infty Y_{2m+1}(t)-t{\bf T}\rr _\BT +|t| 
\|\bT\|_\BT \|e^{tS}-1\|.$$
It is obvious  that 
$$\limt\frac{1}{|t|} |t| \|\bT\|_\BT\|e^{tS}-1\|=0.$$
We have 
$$\frac{1}{|t|}\ll \sum_{m=0}^\infty Y_{2m+1}(t)-t{\bf T}\rr_\BT \leq 
\frac{1}{|t|} 
\|Y_1(t)-t{\bf T}\|_\BT +
\frac{1}{|t|} 
\ll \sum_{m=1}^\infty Y_{2m+1}(t)\rr_\BT .$$
Since 
$$\|e^{t{\bf S}}{\bf T} e^{-t{\bf S}}-{\bf T}\|_\BT \leq 
\|{\bf T}\|_\BT  
(\|e^{-tS}-1\|+\|e^{t S}-1\|),$$
$\|e^{t{\bf S}}{\bf T} e^{-t{\bf S}}-{\bf T}\|_\BT $ is continuous at $t=0$. 
Then 
we have 
\be 
\limt \frac{1}{|t|}
\|Y_1(t)-t{\bf T}\|_\BT 
&=& 
\limt \frac{1}{|t|}
\|\int_0^t (e^{t_1 {\bf S}} {\bf T} e^{-t_1{\bf S}}-{\bf T}) dt_1\|_\BT \\
&\leq &
\limt 
\frac{1}{|t|} \int_0^t  \|e^{t_1 {\bf S}} {\bf T} e^{-t_1{\bf S}}-{\bf T}\|_\BT    dt_1
=0.
\ee 
Moreover  by \kak{ynt2},  
\be\limt\frac{1}{|t|} 
\ll \sum_{m=1}^\infty Y_{2m+1}(t)\rr_\BT 
&\leq& \limt \frac{1}{|t|}\sum_{m=1}^\infty 
\frac{\|T\|^{2m}}{(2m+1)!} |t|^{2m+1} \|{\bf T}\|_\BT \\
&=& \limt  \|{\bf T}\|_\BT  \frac{\sinh (|t|\|T\|)-|t|\|T\|}{|t|\|T\|}=0.
\ee 
Hence we conclude  \kak{SH}. In the case of $i=1,j=2$, 
the lemma is similarly proven. 
\qed

\bl{L2}
$K_t$ is differentiable in $t$ in $B_2$ with 
\eq{l1}
\frac{d}{dt} K_t=K'_t:=(\dA e^{t\dA})_{21}(e^{t\dA})_{11}\f-
(e^{t\dA})_{21}(e^{t\dA})_{11}\f(\dA e^{t\dA})_{11}(e^{t\dA})_{11}\f.
\en 
\el
\proof 
By Lemma \ref{L1} we see that 
\be 
&& \frac{d}{dt} (\ea)_{21}=(\dA\ea)_{21}\ \ in\  B_2,\\
&& 
\frac{d}{dt}(\ea)_{11}\f=(\ea)_{11}\f(\dA\ea)_{11}(\ea)_{11}\f\ \ in \ B.
\ee
Then by Lemma \ref{L0} we obtain \kak{l1}.
\qed

\bl{L4} 
Let $\Psi\in \ffff$. 
Then $(\XY)^n\Psi$ is strongly differentiable in $t$ with 
$$ \frac{d}{dt}(\XY)^n\Psi=n \XX(\XY)^{n-1}\Psi.$$
\el
\proof 
Using Lemma \ref{san},  we have
$$\frac{1}{\epsilon}
\lk 
\dd{K_{t+\epsilon}}-\dd{K_t}
\rk \Psi-\dd{K_t'}\Psi
=
\dd{
\frac{1}{\epsilon}(K_{t+\epsilon}-K_t)-K_t'
}\Psi.$$
Then by Lemmas \ref{funda3} and \ref{L2}, 
we have 
$$
\left 
\| \frac{1}{\epsilon}\lk \dd{K_{t+\epsilon}}-\dd{K_t}\rk \Psi-\dd{K_t'}\Psi\right \|
\leq 
\sqrt 6 
\left 
\|\frac{1}{\epsilon}\lk K_{t+\epsilon}-K_t\rk-K_t'\right\|_2
\|(N+1)\Psi \|
\rightarrow 0$$
as $\epsilon\rightarrow 0$. 
Then 
$$\frac{d}{dt}\XY\Psi=\XX\Psi.$$
We have 
\be
&&
\frac{1}{\epsilon}\lkk (\dd{K_{t+\epsilon}})^n-(\XY)^n\rkk \Psi-n\XX\XY^{n-1}\Psi\\
&&
=\sum_{j=0}^{n-1}(\dd{K_{t+\epsilon}})^j \lkk \frac{1}{\epsilon}(\dd{K_{t+\epsilon}}-\XY)
-\XX\rkk (\XY)^{n-j-1}\Psi\\
&&\hspace{1cm}
+
\sum_{j=0}^{n-1}\sum_{i=0}^{j-1}
(\dd{K_{t+\epsilon}})^i\lk \dd{K_{t+\epsilon}}-\XY \rk \XX(\XY)^{n-i-2}\Psi.
\ee
Hence 
we see that
\bee
&&
\left \|\frac{1}{\epsilon}((\dd{K_{t+\epsilon}})^n-(\XY)^n)\Psi-n\XX\XY^{n-1}\Psi\right\|\nn \\
&&
\leq (\sqrt 6)^n 
\sum_{j=0}^{n-1} (2n-1)!! \|K_{t+\epsilon}\|^j \|K_t\|^{n-j-1}
\left \|\frac{1}{\epsilon}(K_{t+\epsilon}-K_t)-K'_t\right\|_2\|\Psi\|\nn \\
&&
\label{HHHH}
+
(\sqrt 6)^n\sum_{j=0}^{n-1}\sum_{i=0}^{j-1} 
(2n-1)!! \|K_{t+\epsilon}\|^i_2\|K_t'\|_2\|K_t\|^{n-i-2}\|K_{t+\epsilon}-K_t\|_2\|\Psi\|.
\eee
As $\epsilon\rightarrow 0$, 
the right-hand side of \kak{HHHH} goes to zero. Then the lemma follows.
\qed

\bl{L5}
Let $\Psi\in\fff$. Then 
$(\Psi, U_t\Omega)$ is differentiable in $t$.
\el
\proof 
We have 
$$(\Psi, U_t\Omega)={\rm det}(1-K_t^\ast K_t)^{1/4}(\Psi, e^{-\half \dd{K_t}}\Omega).$$
We shall show that 
${\rm det}(1-K_t^\ast K_t)^{1/4}$ and $(\Psi, e^{-\half \dd{K_t}}\Omega)$ are differentiable 
in $t$. 
Note that 
$${\rm det}(1-K_t^\ast K_t)^{-\han}=
\sumn\lk\frac{1}{n!}\lk-\half\rk^n\rk^2\|(\XY)^n\Omega\|^2.$$
Since 
$$\frac{d}{dt}
\|(\XY)^n\Omega\|^2=2\Re ((\XY)^n\Omega, n\XX(\XY)^{n-1}\Omega)$$
by Lemma \ref{L4},  
and 
$$\left| \frac{d}{dt} \|(\XY)^n\Omega\|^2\right|
\leq 2n (2n-1)\sqrt 6 \|K_t'\|_2\|(\XY)^n\Omega\|\|(\XY)^{n-1}\Omega\|,$$
we obtain 
that 
\be &&
\sumn \left| \lk \frac{1}{n!}\lk-\half\rk^n\rk^2 
\frac{d}{dt}\|(\XY)^n\Omega\|^2\right|\\
&&
\leq 
\sqrt 6 \lk\sumn 
\lk \frac{1}{n!}\lk-\half\rk^n\rk^2 \|(\XY)^n\Omega\|^2
\rk^\han \\
&&\hspace{1cm}\times 
\lk\sumn 
(2n+1)^2 \lk \frac{1}{n!}\lk-\half\rk^n\rk^2 
\|(\XY)^n\Omega\|^2
\rk^{\han} <\infty.
\ee 
Thus 
${\rm det}(1-K_t^\ast K_t)^{-\han}$ is continuously 
differentiable in $t$ with 
$$\frac{d}{dt}{\rm det}
(1-K_t^\ast K_t)^{-\han}=2\Re (e^{-\half\XY}\Omega, -\half \XX e^{-\half\XY}\Omega).$$
In particular ${\rm det}
(1-K_t^\ast K_t)^{1/4}$ is continuously differentiable in $t$. 
Next we estimate $(\Psi, e^{-\half \dd{K_t}}\Omega)$.
We have 
$$(\Psi, e^{-\half \dd{K_t}}\Omega)=\sumn\frac{1}{n!}\lk-\half\rk^n(\Psi^{(2n)}, (\XY)^n\Omega).$$
Since by Lemma \ref{L4}, 
$$\frac{d}{dt}(\Psi^{(2n)}, (\XY)^n\Omega)=(\Psi^{(2n)}, n\XX(\XY)^{n-1}\Omega),$$
and
\be
\left|
\frac{d}{dt}(\Psi^{(2n)}, (\XY)^n\Omega)\right|
&\leq& 
\|\Psi^{(2n)}\| 
n(2n-1)\sqrt 6 \|K_t'\|_2\|(\XY)^{n-1}\Omega\|,
\ee 
it follows that 
\be&& \sum_{n=1}^\infty 
\left| 
\frac{d}{dt}(\Psi^{(2n)}, (\XY)^n\Omega)\right|\\
&& \leq 
\|\Psi\| \sqrt 6 \|K_t'\|_2 
\lk\sumn (2n+1)^2\frac{1}{2^2} 
 \lk \frac{1}{n!}\lk-\half\rk^n\rk^2\|(\XY)^n\Omega\|^2\rk^\han<\infty,
\ee 
which implies that 
$(\Psi, e^{-\half \dd{K_t}}\Omega)$ is continuously differentiable in $t$ with 
$$\frac{d}{dt} 
(\Psi, e^{-\half \dd{K_t}}\Omega)
=(\Psi, -\half \XX e^{-\half \dd{K_t}}\Omega).$$
Hence the lemma follows.
\qed

\section{Proof of Theorems}

\subsection{Proof of Theorem \ref{mmm}} 
\bl{N1}
Let $\dA \in\SS $. Then 
$$\|\D(\dA )\Psi\|\leq \alpha \|(N+1)\Psi\|,$$
where 
$\alpha:=\max\{\sqrt 6\|T\|_2,\|S\|\}$. 
\el
\proof 
We obtain from \kak{ss11}--\kak{ss13} that 
\be 
\|\D(\dA )\Psi\|
&\leq& \half\lk\|\dd T \Psi\|+
\|\aa{\ov T}\Psi\|\rk+\|\dA {\ov S}\Psi\|,\\
&\leq& \half\lk\sqrt6\|T\|_2\|(N+1)\Psi\|+\|T\|_2\|N\Psi\|\rk+\|S\|\|N\Psi\|,\\
&\leq& \max\{\sqrt 6\|T\|_2,\|S\|\}\|(N+1)\Psi\|.
\ee
Thus the lemma follows.
\qed
\bl{N2}
Let $\Psi\in\ffff$. Then,  for $t\in (-1/2\alpha, 1/2\alpha)$,  
$$
\sum_{n=0}^\infty\frac{\|\D(\dA )^n\Psi\|t^n}{n!}<\infty.$$
\el
\proof
It is enough to prove the lemma for  $\Psi\in\fff^{(n)}$.
Note that 
$$\D(\dA)^k\Psi\in \bigoplus_{m=0}^{n+2k} \fff^{(m)}.$$
By Lemma \ref{N1} it follows that 
$$
\|\D(\dA )^k\Psi\|
\leq 
\alpha^k(n+2k-1)(n+2k-3)\cdots (n+1)\|\Psi\|.
$$ 
Hence we have 
$$
\sum_{n=0}^\infty\frac{\|\D(\dA )^n\Psi\|t^n}{n!}\leq 
\sum_{n=0}^\infty \frac{(n+2k-1)(n+2k-3)\cdots (n+1)}{n!} \alpha^k t^k\|\Psi\|.$$
The right-hand side above converges for $t\alpha \in (-1/2, 1/2)$, thus the lemma follows.
\qed

{\it Proof of Theorem \ref{mmm}}\\
By Lemma \ref{san} and the assumption such that $\ov{T^\ast}=T$, $S^\ast=-S$, 
it follows that 
$$(\D_T^\dagger)^\ast=\D_{\ov T},\ \ \ 
N_{\ov S}^\ast=-N_{\ov S}$$ on $\ffff$. In particular 
$D(\dA )$ is a  symmetric 
operator on $\ffff$.  
By Lemma \ref{N2}, we see that $\ffff$ is a set of analytic vectors for $\D(\dA )$. 
Hence the Nelson analytic vector theorem \cite[Theorem X.39]{rs2} yields 
that $\D(\dA)$ is essentially self-adjoint on $\ffff$. 
\qed

\subsection{Proof of Theorem \ref{mm}}
Let $\ad_A^0(B)=B$ and $\ad_A^k(B)=[A,\ad_A^{k-1}(B)]$.
It is well known that 
$$A^nB=\sum_{k=0}^n\cnk \ad_A^k(B)A^{n-k}.$$
\bl{N3}
We have on $\ffff$ 
\be 
&& (1)\ \ad_{i\D(\dA )}^k(a(f))=a((\dA ^k)_{11}f)+\add((\dA ^k)_{21}f),\\
&& (2)\ \ad_{i\D(\dA )}^k(\add(f))=a((\dA ^k)_{12}f)+\add((\dA ^k)_{22}f).
\ee 
\el
\proof
We prove the lemma through  an induction. 
We directly see that 
\be 
&&  [\ia, a(f)]=a(Sf)+\add(Tf),\\
&&  [\ia, \add(f)]=a(\ov T f)+\add(\ov S f).
\ee 
Thus the lemma follows for $k=1$.  
Assume that 
$$\ad_{\ia}^k(\add(f))=a((\dA ^k)_{12}f)+\add((\dA ^k)_{22}f).$$
Then we have 
\be 
\ad_{\ia}^k(\add(f))&=& [\ia,  +a((\dA ^k)_{12}f)+\add((\dA ^k)_{22}f)]\\
&=& a(\ov T (\dA ^k)_{22}f+S(\dA ^k)_{12}f)+
\add(\ov S (\dA ^k)_{22}f+T(\dA ^k)_{12}f)\\
&=& a((\dA ^{k+1})_{12}f)+\add((\dA ^{k+1})_{22}f).
\ee
Thus (1) follows.
(2) is proven in the similar manner. 
\qed

{\it Proof of Theorem \ref{mm}}\\
Let $\Psi,\Phi\in\ffff$. 
We define for $z\in\BC$ 
\be
&& F_1(z)=(\Phi, e^{iz\D(\dA )}\add(f)\Psi),\\
&& F_2(z)=(b_{\ov{z}}(\ov f) \Phi, e^{iz\D(\dA )}\Psi),
\ee
where 
$$b_{\ov{z}}(\ov f) = a((e^{\ov z\dA })_{11}\ov f)+\add((e^{\ov z \dA })_{21} \ov f).$$
Since,  by Lemma \ref{N2},  
$\Psi$ and $\Phi$ are analytic vectors for $\D(\dA )$,   
for $|z|<1/2\alpha$, $F_j(z)$ are analystic. 
We have 
\be 
\frac{d^n F_1(z)}{dz^n}\lceil_{z=0}
&=& (\Phi, (\ia)^n\add(f)\Psi)\\
&=& \sum_{k=0}^n \cnk (\Phi, \ad_{\ia}^k(\add(f))(\ia)^{n-k}\Psi)\\
&=& \sum_{k=0}^n \cnk (\Phi, [a((\dA ^k)_{12}f)+\add((\dA ^k)_{22}f)](\ia)^{n-k} \Psi).
\ee
On the other side 
we see that 
\be
\frac{d^n F_2(z)}{dz^n}\lceil_{z=0}
&=& 
\sum_{k=0}^n \cnk ([a((\dA ^k)_{11} \ov f)+\add((\dA ^k)_{21}\ov f) ] \Phi, 
(\ia)^{n-k}\Psi)\\
&=&\sum_{k=0}^n \cnk (\Phi, [a((\dA ^k)_{12}f)+\add((\dA ^k)_{22}f)] (\ia)^{n-k} \Psi).
\ee
Here we used 
$$\ov{(\dA^k)_{21}}={(\dA^k)_{12}},\ \ \ 
\ov{(\dA^k)_{11}}={(\dA^k)_{22}}.$$
Hence 
we obtain
$$F_1(z)=F_2(z),\ \ \ |z|<1/2\alpha,$$
which implies that 
$$e^{-iz\dA }\Phi\in D(a(\ov f))$$
with 
$$a(\ov f) e^{-iz\dA }\Phi= e^{-iz \dA } b_{\ov z} (\ov f)\Phi,\ \ \ |z|<1/2\alpha.$$ 
In particular 
we have 
$$e^{it\dA } a(f) e^{-it\dA }\Phi=  b_t (f)\Phi$$
for all $t\in\BR$ by the group property. 
The identity 
$$e^{it\dA } \add(f) e^{-it\dA }\Phi=  b_t^\ast (f)\Phi$$
can be also proven similarly. 
Then the theorem follows.
\qed

\subsection{Proof of Theorem \ref{mM}}
By the uniqueness of the proper canonical transformation, we have 
$$U_t=e^{i\theta(t)} e^{it\dA},$$
with some function $\theta(\cdot):\RR\rightarrow \RR$. 
\bl{H1}
We have $\theta\in C(\RR)$.
\el
\proof 
Note that 
$$e^{i\theta(t)}=\frac{(\Omega, U_t\Omega)}{(\Omega, e^{it\D(A)}\Omega)}.$$
Since $(\Omega, U_t\Omega)$ and $(\Omega, e^{it\D(A)}\Omega)$ are 
continuously differentiable in $t$ by Lemma \ref{L5}, the lemma follows. 
\qed

{\it Proof of Theorem \ref{mM}}\\
Since 
$$\RR\ni {\rm det}(1-K_t^\ast K_t)^{-\han}=
(\Omega, U_t\Omega)=(\Omega, e^{i\theta(t)}e^{it\D(A)}\Omega),$$
we see that 
$$\frac{d}{dt}
(\Omega, e^{i\theta(t)}e^{it\D(A)}\Omega)=i\theta'(t)(\Omega, U_t\Omega)
+(\Omega, i\D(A)U_t\Omega)\in\RR.$$
Moreover 
$(\Omega, U_t\Omega)\in\RR$ employs that 
$$\theta'(t)(\Omega, U_t\Omega)=
-\Im(\Omega, i\D(A)U_t\Omega)=-\Im(\half\dd T\Omega, U_t\Omega).$$
Then we have 
$$\theta'(t)=-\Im \half \frac{(\dd T\Omega, U_t\Omega)}{(\Omega, U_t\Omega)}
=-\half \Im(\dd T \Omega, e^{-\half \dd{K_t}}\Omega)=\frac{1}{4}\Im (\dd T\Omega, \dd{K_t}\Omega).$$
It can be directly  proven that 
$$(\Omega, [\aa{T^\ast}, \dd{K_t}]\Omega)=2{\rm Tr}(T^\ast {K_t}).$$
Thus we have 
$$\theta'(t)=\half\Im {\rm Tr}(T^\ast {K_t}).$$
Hence we conclude that 
$$\theta(t)=\int_0^t \half\Im {\rm Tr}(T^\ast {K_s}) ds,$$
and then  the theorem follows.
\qed

\section{Examples}
\subsection{Diagonal cases} 
\bl{345}
Let  $\dA =\Sd\in\SS $. 
Then 
\eq{H2}
U_t=e^{it\D(A)}.
\en 
\el
\proof 
$\TA(s)\equiv 0$ implies 
\kak{H2} by Corollary \ref{PP}.
\qed
\bt{SS1}
Let $S^\ast=-S$. Then 
\eq{S2}
\wick{\exp\lk - N_{(1-e^{+t\ov S})}\rk }=\exp\lk tN_{\ov S}\rk.
\en 
\et
\proof
Let 
$\dA=\Sd$. 
 Since 
\begin{eqnarray*}
U_t
&=& \wick{\exp\lk - N_{(1-e^{+t\ov S})}\rk }\\
\exp\lk it\D(A)\rk 
&=& \exp\lk tN_{\ov S}\rk,
\end{eqnarray*}
the lemma follows from Lemma \ref{345}.\footnote{
\kak{S2}  is  shown in \cite[Corollary 5.4]{rui}.}
\qed
In particular setting 
$S=i 1$ in Lemma \ref{SS1},\footnote{$e^{itN}$ can be regarded as the Fourier trasformation on $\fff$. See \cite{se}.} 
we have 
\eq{Se}
\wick{\exp\lk -(1-e^{it}) N\rk}\Psi =\exp\lk it N\rk\Psi .
\en 
Since 
$\Psi\in\ffff$ is an analytic vector for $\exp\lk it N\rk$, 
\kak{Se} can be extended to $t\rightarrow -i\beta$, $\beta\in\RR$, 
i.e., we have for $\Psi\in\ffff$, 
$$
\wick{\exp\left[ (e^\beta-1) N \right] }\Psi =\exp\lk \beta N\rk\Psi.
$$  
In particular we obtain for $\Psi\in \ffff$, 
\eq{see}
\wick{\exp\lk t N \rk }\Psi =\exp\lk \log (t+1)\   N\rk\Psi,\ \ \ t\geq 0.
\en 
\subsection{Integral formulae}
Let  $\fff=\fff(L^2(\RR))$.
Let 
$a(k)$ be  the kernel of $a(f)$ defined by 
$$\lk a(k) \Psi\rk^{(n)}(k_1,\cdots, k_n)=
\sqrt{n+1}\int \Psi^{(n+1)}(k,k_1,\cdots,k_n)dk_1\cdots dk_n.$$ 
It is well known that for $\Psi\in \ffff$, 
$$\int_{\RR^n}\|a(k)\Psi\|^2 dk=\| N^\han \Psi\|^2.$$
This formula can be extended as follows.
\bc{TT}
Let  $\Psi\in\ffff$ and $t\geq0$. Then 
\begin{eqnarray*}
\sum_{n=0}^\infty \frac{t^n}{n!}\int_{\RR^n} \|a(k_1)\cdots a(k_n)\Psi\|^2 dk_1
\cdots dk_n=
\|e^{(\han) \log(t+1)\ N}\Psi\|^2.
\end{eqnarray*}
\ec
\proof
We have 
\begin{eqnarray*}
\sum_{n=0}^\infty \frac{t^n}{n!}\int_{\RR^n} \|a(k_1)\cdots a(k_n)\Psi\|^2 dk_1
\cdots dk_n=(\Psi, \wick{e^{tN}}\Psi)
\end{eqnarray*}
Then 
from \kak{see} the lemma follows.
\qed

\subsection{Commutative  cases}
\bt{comuta}
Let $\dA=\ST\in\SS$ and 
$[S, T]=0$. 
Then 
\bee
&& 
\exp\lk it\lkk \frac{i}{2}(\dd T-\aa{\ov T})-i\da{\ov S}\rkk\rk \nn\\
&&
\eqq{U2}
=C 
\exp\lk -\half\D^\dagger_{\tanh tT}\rk
\wick{\exp\lk - N_{(1-e^{+t\ov S}\cosh\f tT)}\rk }  
\exp\lk +\half \D_{\tanh tT}\rk,
\eee 
where 
\be
&&C:=e^{-i \int_0^t \half \Im {\rm Tr}(T^\ast \tanh sT) ds} {\rm det}(1-|\tanh tT|^2)^{1/4},\\
&&|\tanh tT|^2=(\tanh tT)^\ast \tan htT.
\ee
 \et
\proof
It follows that 
$$\left[ \mat {S} {0} {0}{S}, 
\mat {0}{T}{T}{0}\right]=0.$$
Hence 
we see that 
\bee \exp(tA)
&=& \exp\lk{t\mat {S} {0} {0}{ S}}\rk\exp\lk {t\mat
 {0}{T}{T}{0}}\rk\nn\\
\eqq{U3}
&=& \mat{e^{tS}\cosh tT}{e^{tS}\sinh tT}{e^{tS}\sinh tT}{e^{tS}\cosh tT}.
\eee 
Since 
\be 
&& \TT\OO\f=\OO\f\TT=\tanh tT\\
&& 
1-\T{((e^{t\dA })_{11}\f)}=1-\T{(e^{-tS}\cosh\f tT)}=
1-e^{+t\ov S}\cosh\f tT, 
\ee 
\kak{U2} follows from Theorem \ref{mm}.
\qed

\bc{comuta2}
Assume the same assumptions as in Theorem \ref{comuta}. 
Then 
$$(\Omega, 
e^{ it\lkk \frac{i}{2}(\dd T-\aa{\ov T})-i\da{\ov S}\rkk}\Omega)
=e^{-i \int_0^t \half \Im {\rm Tr}(T^\ast \tanh sT) ds} {\rm det}(1-|\tanh tT|^2)^{1/4}.$$
\ec

\appendix
\section{Appendix}
We demonstrate a direct proof of the following theorem in this Appendix.
\bt{NHK}
Let $\dA=\ST\in\SS$. 
Then for $\Psi\in\ffff$, 
$$s-\limt \frac{U_t-1}{t}\Psi=i\D(\dA)\Psi.$$
\et
Before going to a proof of Theorem \ref{NHK}, we show several lemmas.
 For notational convenience we set 
$$\X:=-\half \dd{\K}, \ \ \ 
\Y=-\half \dd{T}.$$

\bl{Y2}
We have 
$$\limt \Tr\left[ \frac{1}{t^2}\log (1-\K^\ast \K)\rkkk=
-\|T\|_2^2.$$
\el
\proof 
We shall prove 
$$\limt \Tr
\left[ \frac{1}{t^2}\log (1-\K^\ast \K)+T^\ast T\rkkk=0.$$
Note that $\|\K\|<1$. Then 
we have 
$$\log (1-\K^\ast \K)=-\sum_{n=1}^\infty \frac{1}{n}(\K^\ast \K)^n$$
in the operator norm,  from which it follows that 
\be 
\Tr\left[ \frac{1}{t^2}\log (1-\K^\ast \K)+T^\ast T\right]
&=& 
\Tr\lkkk T^\ast T-\frac {1}{t^2}\K^\ast \K\rkkk -
\Tr\lkkk \frac{1}{t^2}\sum_{n=2}^\infty (\K^\ast \K)^n \rkkk\\
&=& \lk \|T\|_2^2-\ll \KK\rr_2^2\rk-
\Tr\lkkk \frac{1}{t^2}\sum_{n=2}^\infty (\K^\ast \K)^n\rkkk.
\ee 
By Lemma \ref{L2}, $K_t/t\rightarrow T$ in $B_2$ as $t\rightarrow 0$. Then 
$$\limt \lk \|T\|_2^2-\ll\KK\rr_2^2\rk =0.$$
Let $\{e_j\}$ be a complete orthonormal system of $\hhh$. 
We have 
\be 
| 
\Tr\lkkk \frac{1}{t^2}\sum_{n=2}^\infty \frac{1}{n} 
(\K^\ast \K)^n\rkkk 
| 
&=& | \sum_{j=1}^\infty (e_j, \frac{1}{t^2}\sum_{n=2}^\infty (\K^\ast \K)^n e_j)|
\\
&\leq & 
\sum_{j=1}^\infty \frac{1}{t^2} \sum_{n=2}^\infty \frac{1}{n} 
| (\K e_j,  \K (\K^\ast \K)^{n-2} \K^\ast \K e_j)|\\
&\leq & 
\sum_{j=1}^\infty \frac{1}{t^2} \sum_{n=2}^\infty\frac{1}{n}  
\|\K e_j\|^2 \|\K\|^{2n-2}\\
&=& 
\sum_{j=1}^\infty 
\frac{\|\K e_j\|^2 }{t^2} 
\lk 
\frac{-\log(1-\|\K\|^2)-\|\K\|^2}{\|\K\|^2}\rk\\
&=& 
\ll \KK \rr _2^2 
\lk \frac{-\log(1-\|\K\|^2)-\|\K\|^2}{\|\K\|^2}\rk.
\ee 
Since 
$\|\K/t\|_2^2\rightarrow \|T\|_2^2$ as $t\rightarrow 0$ 
and 
$\|\K\|\rightarrow 0$ as $t\rightarrow 0$, 
we obtain that 
$$\limt \ll \KK\rr_2^2 
\lk \frac{-\log(1-\|\K\|^2)-\|\K\|^2}{\|\K\|^2}\rk =0.$$
Hence the lemma follows.
\qed

\bl{infinite}
We have 
$$\lim_{t\rightarrow 0}\frac{1}{t^2}\lkk 
{\rm det}(1-\K^\ast \K) ^{1/4}-1\rkk
=-\frac{1}{4}\|T\|_2^2.$$
\el
\proof 
Since $\det\lceil_{t=0}=1$, it follows that 
\be
\limt \frac{1}{t^2}\lkk \det-1\rkk 
&=& \dt{\det}\\
&=& 
\dt{\log \det}.
\ee 
Note that 
$$\log \det =\frac{1}{4} \Tr\log (1-\K^\ast \K).$$
Then we have by Lemma \ref{Y2} 
$$
\limt \frac{1}{t^2}\lkk \det-1\rkk 
=
\frac{1}{4} \limt \frac{1}{t^2} \Tr \log \lk 1-\K^\ast \K \rk
=-\frac{1}{4} \| T\|_2^2.
$$
The proof is complete.
\qed

\bl{K1}
We have 
$$\limt\ll\tu \Omega\rr=
\|\Y\Omega\|.$$
\el
\proof 
We have 
$$\limt\ll\tu \Omega\rr^2=
\limt \frac{2}{t^2}
\lkk 1-\Re (U_t\Omega, \Omega)\rkk.$$
Note that 
$$(U_t\Omega, \Omega)=\det.$$ 
Thus by Lemma \ref{infinite} we have 
$$\limt \frac{2}{t^2}(1-\det)=\half \|T\|_2^2=\|\Y\Omega\|^2.$$
Thus the  lemma follows.
\qed

\bl{111}
Let $k\geq 1$. Then 
$$\limt \frac{1}{t}\|N^kU_t\Omega\|=\|N^k \Y\Omega\|.$$
\el
\proof 
We note that 
$$\frac{1}{t^2}\|N^kU_t\Omega\|^2=
\sum_{n=1}^\infty \frac{1}{(n!)^2}(2n)^{2k}
\ll\frac{(\X)^n}{t}\Omega\rr^2.$$
By Lemma \ref{L4} we have 
$$\limt \ll\frac{(\X)^n}{t}\Omega\rr^2=\lkk\begin{array}{ll}
 \|\Y\Omega\|^2,&n=1,\\
0,&n\geq 2,\end{array}\right.$$
and 
$\limt\|(\X)^n\Omega\|^2/t^2\rightarrow 0$ uniformly in $n\geq 2$. 
Then by the Lebesgue dominated convergence theorem yields that 
\be
&&\limt \frac{1}{t^2}\|N^kU_t\Omega\|^2
=\limt 
\sum_{n=1}^\infty \frac{1}{(n!)^2}(2n)^{2k}
\ll\frac{(\X)^n}{t}\Omega\rr^2\\
&&
=\sum_{n=1}^\infty \frac{1}{(n!)^2}(2n)^{2k}
\limt
 \ll\frac{(\X)^n}{t}\Omega\rr^2=2^{2k}\|\Y\Omega\|^2=\|N^k\Y\Omega\|^2.
\ee 
Thus the lemma follows.
\qed

\bl{222}
We have 
$$s-\dtt {N^k U_t \Omega} =N^k  \Y \Omega.$$
\el
\proof 
(In the case of $k=0$) 

We see that 
\eq{cc}
\ll \tu \Omega-\Y\Omega\rr ^2=
\ll \tu \Omega\rr^2-2\Re\lk \tu \Omega,\Y\Omega\rk+\ll\Y\Omega\rr^2.
\en 
From Lemma \ref{L5} it follows that 
$$\limt \lk \tu \Omega, \Y\Omega\rk=\|\Y\Omega\|^2,$$
and from Lemma  \ref{K1} 
$$\limt\ll\tu \Omega\rr^2=
\|\Y\Omega\|^2.$$
Thus 
$$ \limt \ll \tu \Omega-\Y\Omega\rr =0$$ 
holds and  the  lemma follows for $k=0$.

(In the case of $k\geq 1$) 

We have 
\be && 
\ll N^k \tu \Omega -N^k\Y \Omega\rr^2 \\
&& =
\ll N^k \tu \Omega\rr ^2-2\Re \lk \tu \Omega, N^{2k} \Y \Omega\rk+\ll N^k \Y
 \Omega\rr^2.
\ee 
From Lemma \ref{L5} it follows that 
$$\limt \lk \tu \Omega, N^{2k}\Y \Omega\rk=\|N^k \Y \Omega\|^2,$$
and from Lemma \ref{111} 
$$\limt \frac{1}{t^2}
\ll N^k \tu \Omega\rr^2
=\limt \frac{1}{t^2}\|N^k U_t\Omega\|^2
=\|N^k \Y\Omega\|^2.$$
Thus 
$$\limt \ll N^k \tu \Omega -N^k\Y \Omega\rr=0$$
follows. Hence the proof is complete. 
\qed

{\it Proof of Theorem \ref{NHK}}\\
Let $\Phi=\add(f_1)\cdots \add(f_n)\Omega$. 
Then 
$$\tu \Phi=I(t)+II(t),$$
where 
\be &&
I(t)=
\sum_{j=1}^{n}\bb(f_1)\cdots \bb(f_{j-1})
 \frac{1}{t}\lk \bb(f_j)-\add(f_j)\rk
 \add(f_{j+1})\cdots \add(f_n)\Omega,\\
&&
II(t)=\bb(f_1)\cdots \bb(f_n)\lk \tu \rk \Omega,\\
&& 
\bb(f_j)=\add((e^{t\dA })_{22} f_j) + 
a((e^{t\dA })_{12} f_j).
\ee
Note that 
\be
&& s-\limt (e^{t\dA })_{22} f_j=f_j,\\
&& s-\limt (e^{t\dA })_{12} f_j=0.
\ee 
Moreover 
\be
&& \frac{1}{t}\lk \bb(f_j)-\add(f_j)\rk
=  \add(\frac{1}{t}(e^{t\dA }-E)_{22} f_j) + 
a(\frac{1}{t}(e^{t\dA })_{12} f_j),\\
&& s-\limt \frac{1}{t}(e^{t\dA }-E)_{22} f_j= \ov S f_j,\\
&& s-\limt \frac{1}{t}(e^{t\dA })_{12} f_j=  s-\limt \frac{1}{t}(e^{t\dA }-E)_{12}
 f_j=\ov T f_j.
\ee 
Hence we obtain that 
\bee
s-\limt I(t) &=& \sum_{j=1}^n\add(f_1)\cdots\lk a(\ov T f_j)+\add(\ov S f_j)\rk\cdots
 \add(f_n)\Omega\nn\\
\label{HH1}
&=& \lk\half \D_{\ov T}+N_{\ov S} \rk \add(f_1)\cdots \add(f_n)\Omega.
\eee
Here we used the assumption $\ov {T^\ast}=T$. 
Next we shall estimate $II(t)$. 
We have 
\be 
&& \ll\bb(f_1)\cdots \bb(f_n) \lkk\tu-(-\half\D_T^\dagger) \rkk  \Omega\rr\\
&& \leq C\ll (N+1)^{n/2}  \lkk\tu-(-\half\D_T^\dagger)\rkk \Omega\rr
\ee 
with some constant $C$ independent of $t$. 
By Lemma \ref{222} it follows that 
$$\limt \ll (N+1)^{n/2}\lkk\tu-(-\half\D_T^\dagger) \rkk \Omega\rr=0.$$
Hence we obtain that 
\be s-\limt II(t) 
&=& s-\limt 
\bb(f_1)\cdots \bb(f_n) (-\half\D_T^\dagger) \Omega\\
&=& 
\add(f_1)\cdots\add(f_n) (-\half\D_T^\dagger) \Omega\\
\eqq{HH2}
&=& -\half\D_T^\dagger  \add(f_1)\cdots\add(f_n)  \Omega.
\ee 
By \kak{HH1} and \kak{HH2} the theorem follows.
\qed

{\it Acknowledgements.} 
{\footnotesize We thank H. Araki and 
S. N. M. Ruijsenaars for helpful comments. 
F.H. thanks  Grant-in-Aid  for Science Research (C) 1554019
from MEXT.
}

\end{document}